\documentclass{article}
\usepackage{spconf,amsmath,graphicx,hyperref}
\usepackage{algorithm}
\usepackage{algorithmic}
\usepackage{amsfonts}
\usepackage{tikz}
\usepackage{multirow}
\usepackage{makecell}
\usetikzlibrary{arrows.meta}
\usepackage{booktabs}
\title{PADP: Perceptual Audio Data Perturbation for Probing Perception Awareness in Audio Quality Models}
\name{
  Guanxin Jiang$^{1}$,
  Andreas Brendel$^{2}$\sthanks{Andreas Brendel has been supported by the Free State of Bavaria by the DSgenAI project.},
  Pablo M. Delgado$^{2}$,
  Jürgen Herre$^{1,2}$
}

\address{
  $^{1}$International Audio Laboratories Erlangen\sthanks{A joint institution of the Friedrich-Alexander Universität Erlangen-Nürnberg (FAU) and Fraunhofer IIS }, Germany \\
  $^{2}$Fraunhofer Institute for Integrated Circuits IIS, Erlangen, Germany
}
\begin{document}
\ninept
\maketitle
\begin{abstract}
This paper presents a collection of audio transforms that introduce perceptually irrelevant distortions and demonstrates their use as perceptual stress tests for audio quality models. We refer to these methods as \textbf{P}erceptual \textbf{A}udio \textbf{D}ata \textbf{P}erturbation (PADP). PADP exploits the insensitivity of the human auditory system to certain fine-grained signal variations to substantially alter the waveform while preserving perceived audio quality and content. The audibility of PADP and the selection of its parameters are evaluated through controlled listening tests, ensuring that the transformations achieve transparent or near-transparent quality for both non-critical and critical items. We further probe the robustness of state-of-the-art (SOTA) perception-motivated objective audio quality models and foundation models. The results reveal a misalignment between model responses and human auditory perception, highlighting the limited perceptual awareness of these models for certain proposed transforms.
\end{abstract}
\begin{keywords}
Perceptual audio quality, psychoacoustics, audio perturbation, objective audio quality metrics, foundation models
\end{keywords}
\section{Introduction}
\label{sec:intro}
Human auditory perception is a complex psychoacoustic and cognitive process that selectively extracts perceptually relevant information rather than faithfully representing the acoustic waveform \cite{ZwickerFastl1999}. The auditory system decomposes sound into frequency-selective channels, or critical bands, within which acoustic energy is integrated and strong perceptual interactions occur \cite{Fletcher1940,Zwicker1961Bark}. The development of psychoacoustics has profoundly influenced perceptual audio coding systems, such as mp3 and AAC \cite{ISO11172-3, ISO13818-7}, through perceptual irrelevancy reduction, as well as perceptually motivated objective quality metrics for evaluating coding distortions. Many objective audio quality metrics incorporate psychoacoustic and auditory perception principles. \cite{9388867} investigates several objective quality tools, including Perceptual Evaluation of Audio Quality (PEAQ) \cite{kabal2002peaq}, 2f-model \cite{8937179}, HAAQI \cite{Kates2016TheHA}, and ViSQOL \cite{chinen2020visqol} on audio coding and source separation. Another family of objective audio quality metrics is learning-based. SCOREQ \cite{Ragano2024SCOREQ} leverages wav2vec 2.0 \cite{Baevski2020Wav2Vec2}, while DeePAQ \cite{Jiang2026DeePAQ} leverages MERT \cite{li2023mert} together with contrastive learning to estimate speech or audio quality. For the aforementioned audio quality metrics, it remains unclear whether audio quality is assessed in a manner consistent with human perception, regardless of whether perceptually relevant characteristics are explicitly modeled through psychoacoustic principles or implicitly learned from the training data. If a robust and precise metric is expected to closely reflect human judgments of audio quality, it should, among other properties, remain insensitive to signal modifications that are perceptually irrelevant to human listeners.

Psychoacoustic principles have been used to constrain adversarial perturbations for Automatic Speech Recognition systems \cite{Schoenherr2019,Qin2019}, but these attacks often rely on masking principles and lack rigorous subjective validation of perceptual transparency. We therefore develop a set of signal modifications that are perceptually irrelevant to human listeners to stress test objective audio quality metrics and foundation models. Our contributions are summarized as follows: 
\begin{enumerate}
    \item We introduce \textbf{PADP} (\textbf{P}erceptual \textbf{A}udio \textbf{D}ata \textbf{P}erturbation), a collection of audio transforms that introduce perceptually irrelevant distortions and can serve as robustness probes for audio quality models. PADP comprises \textbf{APX} (All-Pass Band Phase-Shift Combination), \textbf{SCD} (Scaled Coding Distortion), \textbf{DEL} (Frequency-Selective Delay), and \textbf{DIF} (Decorrelator/Diffusion). 
    \item We show that these transforms substantially alter the waveform while preserving perceived sound quality and content across diverse signal types, as confirmed by subjective listening experiments. 
    \item We demonstrate that such perceptually equivalent transformations can challenge SOTA audio quality assessment systems and foundation models, exposing discrepancies between model responses and human perceptual judgments for perturbations that are irrelevant to listeners.
\end{enumerate} 
\section{Method}
\label{sec:method}
Following the MPEG Surround framework \cite{ISO23003-1, Herre2008mpeg}, we use Quadrature Mirror Filterbanks (QMFs) to decompose the broadband audio signal into equally spaced frequency subbands, which can be grouped to approximately match perceptual frequency regions while enabling near-alias-free reconstruction after modification \cite{BreebaartFaller2007, Schuijers2004LowComplexity, Breebaart2005Parametric}. Given a discrete-time audio frame $\mathbf{x} \!\in \!\mathbb{R}^{M}$, the $F$-band QMF analysis operator $\mathcal{A}_{\mathrm{QMF}}$ processes successive blocks of $N$-samples and maps the frame into $F$ complex-valued subbands

\begin{equation}
\mathbf{X}
=
\mathcal{A}_{\mathrm{QMF}}\!\left(\mathbf{x}\right)
\in \mathbb{C}^{F \times {T}}, \quad \text{with}\quad T=\left\lfloor \frac{M}{N} \right\rfloor,
\label{eq:qmf_analysis}
\end{equation}
where each column of $\mathbf{X}$ corresponds to one analysis block. The resulting $F$ QMF subbands are partitioned into $f$ perceptual bands, with their correspondence specified by a fixed binary mapping matrix $\mathbf{G} \!\in \!\{0,1\}^{F \times f}$. This mapping associates each perceptual band with fewer QMF subbands at low frequencies and progressively more at higher frequencies.

\subsection{All-Pass Band Phase Shift Combination (APX)}
\label{ssec:apx}
APX is motivated by auditory frequency selectivity and the limited sensitivity of human hearing to certain phase modifications within auditory bands, particularly when spectral magnitudes are preserved \cite{ZwickerFastl1999,Fletcher1940,Zwicker1961Bark}. To ensure inaudibility in the low frequencies and reduce the number of free parameters, it groups the perceptual bands using an offset $o\! \in\!\{0, 1, \ldots, \!f-1\}$ and a stride $s\! \in\! \{1, 2, \ldots,\!f-1\}$. The offset excludes the lowest $o$ bands, while the stride groups every $s$ consecutive bands under a shared phase shift independently sampled from a discrete uniform distribution
\begin{equation}
\tilde{\phi}_b \stackrel{\text{i.i.d.}}{\sim} \mathrm{Unif}(\mathcal{A}_\alpha),
\qquad b = 0, 1, \ldots, B-1,
\end{equation}
where $B\!=\!\left\lceil \!\frac{f-o}{s} \!\right\rceil$ and
\begin{equation}
\mathcal{A}_\alpha
=
\left\{
k\alpha : k = 0, 1, \ldots, S-1
\right\},
\qquad
S = \left\lfloor \frac{2\pi}{\alpha} \right\rfloor,
\label{eq:angle_alphabet}
\end{equation}
where $\alpha\!\in\!(0,2\pi)$ is the angular step size. The sampled phases are collected into
\begin{equation}
\tilde{\boldsymbol{\phi}}
=
\left[\tilde{\phi}_0, \tilde{\phi}_1, \ldots, \tilde{\phi}_{B-1}\right]^{\top}
\in \mathcal{A}_\alpha^{B}.
\label{eq:phi_tilde}
\end{equation}
These $B$ phases are expanded to the $f$ perceptual bands by assigning zero to the excluded bands and repeating each phase across its corresponding stride group
\begin{equation}
\phi_i =
\begin{cases}
0, & 0 \le i < o, \\[8pt]
\tilde{\phi}_{\left\lfloor \frac{i-o}{s} \right\rfloor}, & i \ge o,
\end{cases}
\qquad i = 0, 1, \ldots, f-1,
\label{eq:phi_expand}
\end{equation}
which yields the perceptual-band phase vector
\begin{equation}
\boldsymbol{\phi}
=
\left[\phi_0, \phi_1, \ldots, \phi_{f-1}\right]^{\top}
\in \mathbb{R}^{f}.
\label{eq:phi}
\end{equation}
The $f$ perceptual-band phases are then mapped to the $F$ QMF subbands with the binary mapping matrix $\mathbf{G}$ by
\begin{equation}
\boldsymbol{\theta} = \mathbf{G} \boldsymbol{\phi}
\in \mathbb{R}^{F}.
\label{eq:theta}
\end{equation}
Finally, a band-wise phase rotation is applied
\begin{equation}
\mathbf{X}'
=
\mathbf{X}
\odot
\exp\!\left(j \boldsymbol{\theta}\mathbf{1}_T^{\top}\right),
\label{eq:phase_rotation}
\end{equation}
where $\odot$ denotes element-wise multiplication, $\mathbf{1}_T$ is the all ones vector of length $T$, and $\exp(\cdot)$ is applied element-wise. The QMF synthesis operator $\mathcal{S}_{\mathrm{QMF}}$ reconstructs transformed QMF subbands back to time domain
\begin{equation}
\mathbf{x}'
=
\mathcal{S}_{\mathrm{QMF}}\!\left(\mathbf{X}'\right)
\in \mathbb{R}^{M}.
\label{eq:qmf_synthesis}
\end{equation} 

\subsection{Frequency Selective Delay (DEL)}
\label{ssec:del}
DEL is motivated by the frequency-dependent temporal resolution of the auditory system \cite{ZwickerFastl1999}, while the perceptual transparency of the selected delay pattern is established by listening experiments. Imperceptible delays generally increase toward lower frequencies, larger delays are assigned to lower perceptual bands. After QMF analysis in \eqref{eq:qmf_analysis}, a fixed delay template is defined over the $f$ perceptual bands as

\begin{equation}
\boldsymbol{\delta}
=
[\delta_0,\delta_1,\ldots,\delta_{f-1}]^{\top}
\in
\mathbb{N}_{0}^{f},
\label{eq:del_template}
\end{equation}
where $\delta_p$ denotes the delay in QMF samples for perceptual band $p$. In our implementation, delays are applied only to selected bands, e.g., $\delta_0\!=\!4$, \!$\delta_2\!=\!2$, \!$\delta_4\!=\!\delta_6\!=\!\ldots\!=\!\delta_{14}\!=\!1$, with zero delay elsewhere. This delay template is then expanded to the $F$ QMF subbands using the same binary mapping matrix $\mathbf{G}$ as APX

\begin{equation}
\boldsymbol{\tau}
=
\mathbf{G}\boldsymbol{\delta}
=
[\tau_0,\tau_1,\ldots,\tau_{F-1}]^{\top}
\in
\mathbb{N}_{0}^{F},
\label{eq:del_expand}
\end{equation}
where $\tau_k$ denotes the delay assigned to QMF subband $k$.
The delayed coefficient of QMF subband $k$ at frame $t$ is given by

\begin{equation}
X'_{k, t}
=
\begin{cases}
0, & t<\tau_k,\\[4pt]
X_{k, t -\tau_k}, & t\ge\tau_k.
\end{cases}
\label{eq:del_delay}
\end{equation}
Equivalently, let $\mathbf{X}'=[\mathbf{X}'_{0},\mathbf{X}'_{1},\ldots,\mathbf{X}'_{F-1}]^{\top}\in\mathbb{C}^{F\times T}$, 
where $\mathbf{X}'_{k}\!=\![X'_{k,t}]_{t=0}^{T-1}\!\in\!\mathbb{C}^{T}$ denotes the delayed coefficient sequence of the $k$-th QMF subband. The delayed QMF coefficients are then transformed back to the time domain using the QMF synthesis operator $\mathcal{S}_{\mathrm{QMF}}$ in \eqref{eq:qmf_synthesis}.

\subsection{Scaled Coding Distortion (SCD)}
\label{ssec:scd}
SCD builds on perceptual audio coding, which exploits auditory masking to introduce coding errors that can remain perceptually irrelevant at sufficient high quality settings \cite{ISO11172-3,ISO13818-7,9388867}. The reconstruction for an audio signal $\mathbf{x}$ produced by a codec $\mathcal{C}$ at bitrate $r$ is denoted as

\begin{equation}
\mathbf{x}_{r} = \mathcal{C}_{r}(\mathbf{x}),
\end{equation}
where $\mathbf{x}_{r}$ is assumed to be perceptually indistinguishable from $\mathbf{x}$ when $r$ is sufficiently large. 

This distortion can be modeled as perceptually negligible additive noise induced by compression. Based on this formulation, we construct additional perceptually transparent variants via linear interpolation between the original signal and its coded reconstruction with sufficiently small $\lambda$, as verified by listening tests

\begin{equation}
\mathbf{x}' = \mathbf{x} + \lambda (\mathbf{x}_{r} - \mathbf{x})
= (1 - \lambda)\mathbf{x} + \lambda \mathbf{x}_{r},
\label{eq:interpolation_scd}
\end{equation}
where $\lambda \!\geq \!0$ controls the strength of the codec-induced perturbation and a value of 1 reconstructs the original coding result $\mathbf{x}_{r}$.

\subsection{Decorrelator / Diffusion (DIF)}
\label{ssec:dif}
DIF builds on perceptually motivated all-pass decorrelation used in artificial reverberation and spatial audio coding, which reduces signal coherence while largely preserving spectral magnitude and timbre \cite{Schroeder1962NaturalSound,Herre2008mpeg,BreebaartFaller2007}. A decorrelator (``diffusor") generates an output uncorrelated with the input while remaining perceptually similar \cite{Schroeder1962NaturalSound}. The similarity between the left and the right channel is quantified by inter-channel coherence (ICC) in perceptual frequency bands. MPEG Surround uses the following ICC quantization table \cite{Herre2008mpeg}, indexed by $d\!\in\!\{0,\ldots,5\}$, to define candidate target ICC values
\[
\mathbf{f}_{\mathrm{ICC}}
=
\begin{bmatrix}
1 & 0.937 & 0.84118 & 0.60092 & 0.36764 & 0
\end{bmatrix}^{\mathsf{T}},
\]
so that \(f_{\mathrm{ICC},0}\!=\!1\) corresponds to the weakest spatial diffusion and fully correlated channels, and \(f_{\mathrm{ICC},5}\!=\!0\) to the widest perceived spatial image and, hence, to uncorrelated signal channels.

The parameter $\alpha_d$ determines the relative contributions between the direct and decorrelated components and thus determines the amount of spatial diffusion
\begin{equation}
\alpha_d
=
\frac12
\arccos\!\left(f_{\mathrm{ICC},d}\right).
\label{eq:alpha}
\end{equation}
Consequently, $f_{\mathrm{ICC},0}\!=\!1$ yields $\alpha_0\!=\!0$, corresponding to the direct path only, whereas $f_{\mathrm{ICC},5}\!=\!0$ yields $\alpha_5\!=\!\pi/4$, corresponding to the maximum decorrelated contribution. We assume equal left- and right-channel power, i.e., zero inter-channel intensity difference (IID), giving a channel amplitude gain $c\!=\!\frac{1}{\sqrt2}$. The direct and decorrelated mixing gains for the $F$ QMF subbands are

\begin{equation}
h^{\text{direct}}_d
=
c\cos(\alpha_d),
\qquad
h^{\text{diffuse}}_d
=
c\sin(\alpha_d).
\label{eq:mixing_gain}
\end{equation}
$h^{\text{direct}}$ weights the direct component, while $h^{\text{diffuse}}$ controls the decorrelated contribution. Linear mixing of the original and decorrelated signals controls the degree of decorrelation and spatial diffusion while largely preserving perceived timbre. 

A decorrelator $\mathcal R(\cdot)$ is implemented using lattice all-pass filters in the QMF domain as specified in MPEG Surround \cite{Herre2008mpeg, Engdegard2004Synthetic, Herre2005Reference}

\begin{equation}
\mathbf{D}
=
\mathcal{R}(\mathbf{X}) \in \mathbb{C}^{F\times T}.
\label{eq:decorrelator}
\end{equation}
Then, left and right QMF subbands are formed by

\begin{equation}
\mathbf{L}
=
h^{\text{direct}}_d\mathbf{X}
+
h^{\text{diffuse}}_d\mathbf{D},
\quad
\mathbf{R}
=
h^{\text{direct}}_d\mathbf{X}
-
h^{\text{diffuse}}_d\mathbf{D}.
\label{eq:LR_qmf}
\end{equation}
Finally, the left and right QMF representations are transformed back to time domain using the QMF synthesis operator

\begin{equation}
\mathbf{x}'_L
=
\mathcal{S}_{\mathrm{QMF}}(\mathbf{L}),
\qquad
\mathbf{x}'_R
=
\mathcal{S}_{\mathrm{QMF}}(\mathbf{R}).
\label{eq:stereo_synthesis}
\end{equation}

\section{Experiments}
\label{sec:exp}
\subsection{Validation of Inaudibility of PADP}
\label{ssec:valid}
We collect $12$ clean signals from the Open Dataset of Audio Quality (ODAQ) \cite{Torcoli2024ODAQ}, covering average and critical items across diverse content, to assess the audibility of the proposed PADP transforms. Following the MPEG Surround configuration \cite{Herre2008mpeg}, we use $F=64$ QMF filterbanks, $f\!=\!23$ perceptual bands, a frame size of $M\!=\!1024$ and an analysis block of $N\!=\!64$, yielding $T\!=\!16$ blocks per frame. For APX, we adopt an offset of $o\!=\!1$, a stride of $s\!=\!4$, and an angular step of $\alpha\!=\!\pi/2$, resulting in $S\!=\!4$, $B\!=\!6$ and $S^B\!=\!4096$ possible phase-shift combinations. Given this large combination space and the impractical listener burden of evaluating all variants, we randomly sample $12$ combinations without replacement for the listening tests. For DIF, we apply diffuseness values of $d\!=\!1,3,5$ to mono inputs. For SCD, we use mp3 at a bitrate of $r\!=\!80$ kbps with $\lambda \!\in \!\{0.2, 0.6, 0.9, 1.2\}$. Following the design of MUlti Stimulus test with Hidden Reference and Anchor (MUSHRA) \cite{itu_bs1534_3}, we additionally include a $3.5$kHz low-pass filtered anchor and a hidden reference. All investigations, including listening tests and subsequent model evaluations, use on monophonic signals (only the left channel). For DIF processing, the left channel of its stereo output is picked. The loudness level is compensated after conversion to mono.

We conduct two separate MUSHRA tests (one for SCD and one for APX, DEL, and DIF), to limit the number of conditions per session and thus reduce listener fatigue and improve sensitivity to subtle differences. The two tests therefore include six and seven conditions, respectively. Listeners participate using high-quality headphones in a quiet environment. Results are post-screened according to the MUSHRA criteria outlined in \cite{itu_bs1534_3}. Only listeners with reliable and discriminative ratings across both tests are retained (assessors are excluded if they rate the severely impaired anchor above 90 or repeatedly rate the hidden reference substantially below 90). Consequently, nine and eight listeners are retained for the SCD and APX/DEL/DIF tests, respectively. The results are presented and discussed in the following section. % 
\subsection{Robustness of Audio Models under PADPs}
\label{ssec:robust}
To assess the robustness of objective audio quality models to PADPs, we evaluate four perception-motivated models: PEAQ Basic \cite{kabal2002peaq}, the 2f-model \cite{8937179}, ViSQOL v3 \cite{chinen2020visqol}, and HAAQI \cite{Kates2016TheHA}; two learning-based models, SCOREQ \cite{Ragano2024SCOREQ} and DeePAQ \cite{Jiang2026DeePAQ}; and four audio foundation models, MERT \cite{li2023mert} with 95M parameters, MuQ \cite{zhu2025muq}, the audio encoder of CLAP \cite{10095889}, and BASE wav2vec 2.0 \cite{Baevski2020Wav2Vec2}. Although originally developed for music information retrieval, these foundation models have shown generalization to downstream audio quality assessment tasks \cite{Ragano2024SCOREQ,Jiang2026DeePAQ}. Each PADP-transformed signal is evaluated against its time- and level-aligned reference. For SCOREQ, DeePAQ and the foundation models, which do not directly output a conventional quality score for full-references setting, we instead use the Euclidean distance between reference and transformed-signal embeddings. These model outputs are then compared with the average subjective ratings from listening tests.
\section{Discussion and Results}
\label{sec:results}
\subsection{Listening Tests}
\label{ssec:slt}
\begin{figure*}[!t]
\centering
\includegraphics[width=0.93\textwidth]{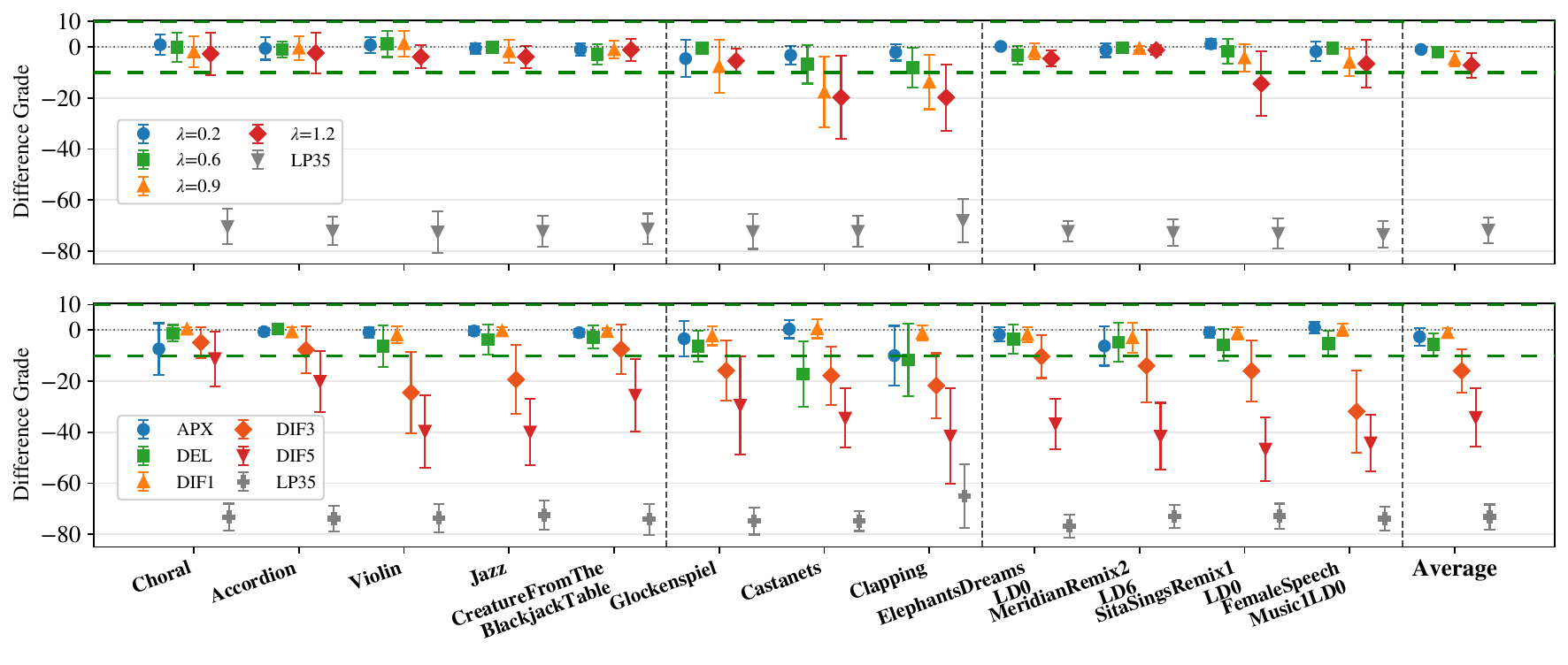}\vspace{-5pt}
\caption{Subjective difference grades for SCD (top) and APX/DEL/DIF (bottom). The green dashed lines at ±10 indicate the adopted transparency threshold.}
\label{fig:subjective_padp}
\end{figure*}

For simplicity of notation, we denote DIF with $d\!=\!1$, $3$, and $5$ as DIF1, DIF3, and DIF5, respectively. Based on the listening results, we adopt a difference grade (i.e. the difference of the grades of the hidden reference and an item) threshold of less than $10$ relative to the reference as the transparency criterion, within which listeners generally struggle to distinguish PADP-transformed signals from the reference. Fig.~\ref{fig:subjective_padp} shows the subjective difference grades for the two listening tests. APX, DIF1 (and potentially DIF2), and SCD with $\lambda \!\leq \!0.6$ consistently exhibit transparent quality across all signal types. Distortions are slightly more pronounced for critical items, with DEL and SCD at $\lambda\!=\!0.9$ occasionally exceeding the transparency range. SCD with $\lambda \!\geq \!1.2$ and DIF3-5 introduce slightly audible distortions while preserving signal content. Thus, for quality-sensitive applications such as audio quality evaluation and neural audio coding, APX, DIF1-2, and SCD with $\lambda \!\leq \!0.6$ offer robust transparency, whereas less stringent applications can additionally employ DEL, DIF3, and SCD with $\lambda\! \geq \!0.9$.

\subsection{Robustness of Audio Quality Metrics}
\label{ssec:poaqm}
\begin{figure}[!t]
\centering
\includegraphics[width=0.93\columnwidth]{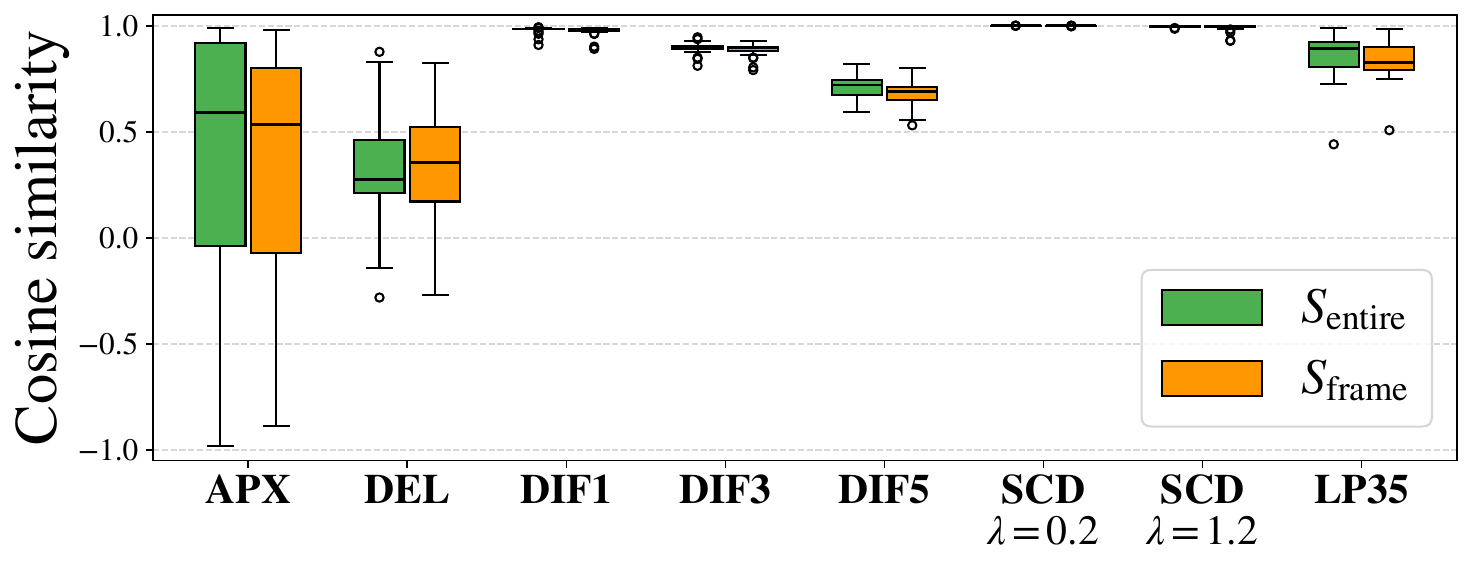} 
\caption{Cosine similarity between PADP-transformed signals and their references, with the low-pass anchor included for comparison.}
\label{fig:similarity}
\end{figure}

We first measure waveform variation using cosine similarity between each PADP-transformed signal and its corresponding reference. Let $\mathbf{x}'$ and $\mathbf{x}$ denote the transformed and reference signals, respectively, each consisting of $T$ frames, with $\mathbf{x}'_t$ and $\mathbf{x}_t$ denoting their $t$-th frames. We consider signal- and frame-level similarity
$s_{\mathrm{entire}} \!= \!\mathrm{cos}(\mathbf{x}',\!\mathbf{x}),
s_{\mathrm{frame}} \!= \!\frac{1}{T}\!\sum_{t=0}^{T-1}\!\mathrm{cos}(\mathbf{x}'_t,\!\mathbf{x}_t)$, where $\mathrm{cos}(\cdot,\cdot)$ denotes the cosine similarity between two vectors. SCD with $\lambda\!=\!0.6$ and $0.9$ are omitted due to negligible differences. As shown in Fig.~\ref{fig:similarity}, APX, DEL, and DIF5 substantially alter the waveform, whereas SCD introduces smaller changes. However, waveform similarity has limited correspondence with human auditory perception: APX, DEL, and DIF5 are perceptually less audible and disturbing than LP35, despite more evident waveform changes. This further motivates PADP as a diagnostic tool for probing the perceptual robustness of models beyond waveform similarity, particularly for applications involving waveform-domain reconstruction.

\begin{figure}[!t]
\centering
\includegraphics[width=0.93\columnwidth]{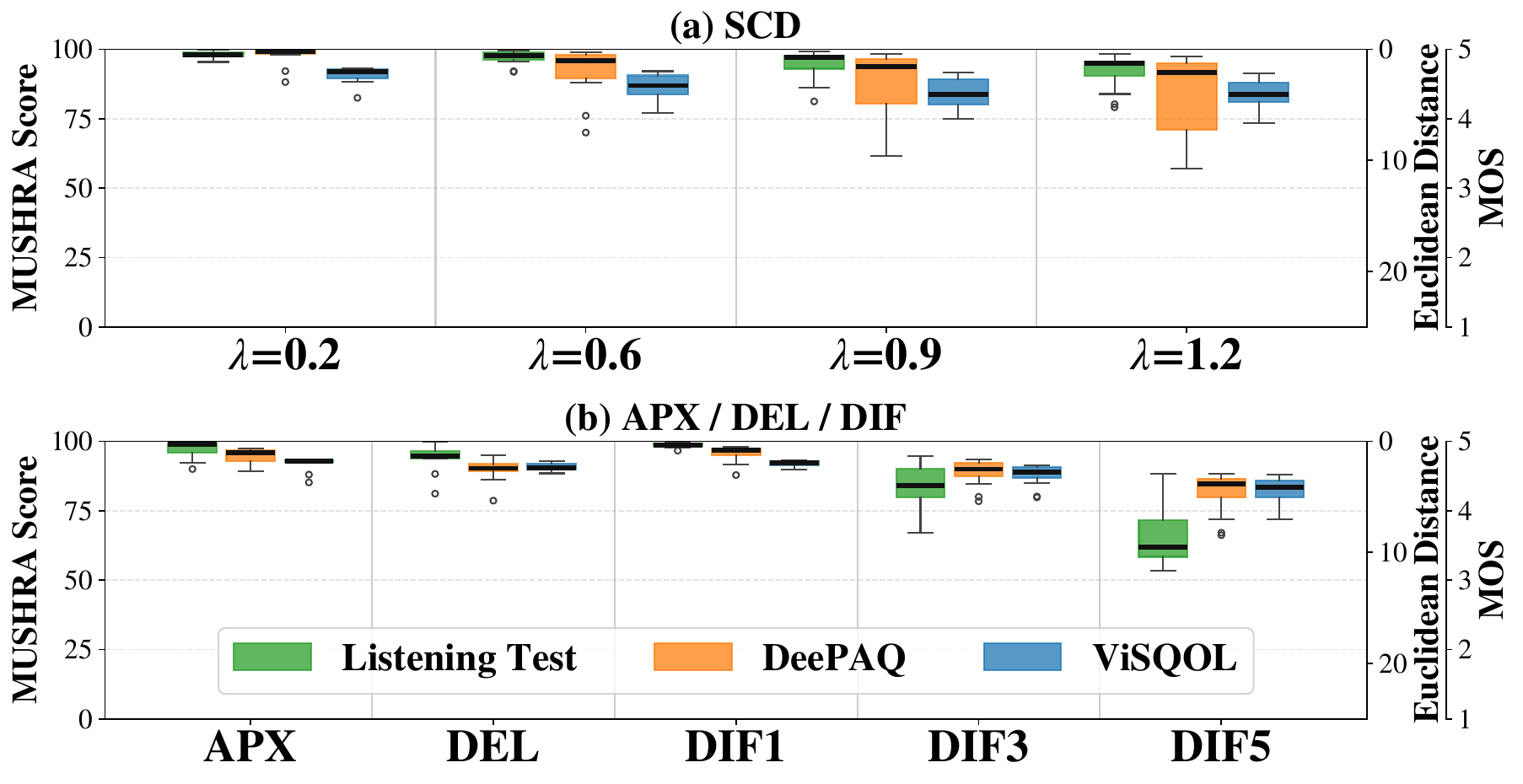} 
\caption{Comparison of DeePAQ and ViSQOL predictions and subjective ratings across signals and PADP transformation conditions. }
\label{fig:padp_deepaq}
\end{figure}

Table~\ref{tab:model_correlation} summarizes the Spearman rank correlation coefficients (SRCCs) between all evaluated model estimates and the mean subjective ratings across all PADP conditions, excluding references and anchors. We omit Pearson correlation because ceiling effects can compress subjective rating variance and make linear correlation less representative. For SCOREQ, DeePAQ, and the foundation models, SRCC is computed between embedding distance and subjective ratings, yielding negative correlations. DeePAQ achieves the highest absolute SRCC ($0.74$), while ViSQOL v3 achieves the lowest ($0.38$), with the remaining models ranging from $0.60$ to $0.70$. Fig.~\ref{fig:padp_deepaq} shows that DeePAQ and ViSQOL roughly preserves the relative ranking of PADP conditions. ViSQOL shows larger deviations from the average subjective ratings, while DeePAQ exhibits wider prediction distributions for SCD and narrower ones for DIF3-5. This indicates that DeePAQ is slightly more sensitive to SCD and less sensitive to DIF than human listeners. Overall, objective audio quality models exhibit varying degrees of performance degradation under the proposed transforms, whereas foundation models, which are evaluated on tasks beyond their original design objectives, show limited accuracy under PADPs. These differences highlight a substantial gap between how the evaluated models interpret such transformations and how they are perceived by human listeners.
\begin{table}[!t]
    \centering
    \small
    \setlength{\tabcolsep}{4pt}    
    \begin{tabular}{c|lc}
        \toprule
        \textbf{Model Type} & \textbf{Models / Metrics} & \textbf{SRCC} \\
        \toprule
        
        \multirow{4}{*}{\makecell[c]{\textbf{Objective}\\\textbf{Quality}\\\textbf{Metrics}}}
        & PEAQ ODG  & 0.63 \\ 
        & 2f-model          & 0.61 \\ 
        & ViSQOL v3   & \textbf{\textcolor{red}{0.38}} \\ 
        & HAAQI       & 0.69 \\
        \midrule
        
        \multirow{2}{*}{\makecell[c]{\textbf{Contrastive}\\\textbf{Learning}}}
        & SCOREQ      & -0.60 \\ 
        & DeePAQ      & \textbf{-0.74} \\
        \midrule
        
        \multirow{4}{*}{\makecell[c]{\textbf{Foundation}\\\textbf{Models}}}
        & MERT        & -0.68 \\ 
        & MuQ         & -0.66 \\
        & CLAP        & -0.63 \\ 
        & wav2vec 2.0 & -0.60 \\
        \bottomrule       
    \end{tabular}
     \caption{Spearman correlations (SRCC) between audio quality estimates and subjective ratings.}
    \label{tab:model_correlation}
\end{table}
\section{Conclusion}
\label{sec:conc}
This paper presents PADP, a collection of audio transforms (i.e., APX, DEL, DIF, and SCD) that introduce perceptually irrelevant distortions. Listening tests verify their transparency and identify suitable parameter ranges: APX, DIF1–2, and SCD with $\lambda \!\leq\! 0.6$ remain transparent across all signal types, while DEL, DIF3, and SCD with $\lambda \!\leq \!0.9$ remain applicable when critical signals are excluded. Applying PADP to objective audio quality metrics and foundation models reveals their susceptibility to these transformations and suggesting PADP as a potential data augmentation technique for improving perceptual awareness and robustness.

%\vfill
%\pagebreak

\clearpage
\bibliographystyle{IEEEbib}
\bibliography{strings,refs}

\end{document}